\documentclass[acus]{JAC2000_1}
\usepackage{graphicx}

\begin{document}
\title{\flushright{T20}\\[15pt] \centering THE INVERSE PROBLEM:\\
EXTRACTING TIME-LIKE FROM SPACE-LIKE DATA}

\author{R. Baldini, E Pasqualucci, INFN Laboratori Nazionali di Frascati, Italy\\
S. Dubni\v{c}ka, Ist. of Physics, Slovak Acad. of Sciences, Bratislava, Slovak Republic\\
P. Gauzzi, Dipartimento di Fisica dell'Universit\`a and INFN, Roma, Italy\\
S. Pacetti, Dipartimento di Fisica dell'Universit\`a and INFN, Perugia, Italy\\
Y. Srivastava, Dipartimento di Fisica dell'Universit\`a and INFN, Perugia, Italy and\\
Physics Departement, Northeastern University, Boston, MASS, USA}

\maketitle

\begin{abstract}
A practical strategy is presented and successfully implemented to determine  
form factors in the time-like but unphysical (below threshold) region using
dispersion relations, in a model independent way without any bias towards
expected resonances. Space and time-like data have been employed along with
a regularization scheme to unfold and solve the integral equations. 
Remarkably, resonance structures with peaks for the $\rho(770)$, 
$\rho'(1600)$ and a structure near the $N\bar N$ threshold are 
automatically generated. The $\Phi$ peak is invisible thus refuting 
suggestions about any sizeable $s\bar s$ content in the nucleon.
\end{abstract}

\section{THE ``IN PRINCIPLE'' METHOD}

Consider a dispersion relation (DR) for a (generic) normalized nucleon
form factor (FF) $G(s)$ with a subtraction at $s\ =\ 0$ and $t<0$
$$
\ln G(t) = {{t\sqrt{s_o - t}}\over{\pi}}
\int_{s_o}^\infty{{\ln|G(s)|ds}\over{s(s - t)\sqrt{s - s_o}}}\eqno(1)
$$
For $s>s_o$ (the lowest mesonic threshold, $4m_\pi^2$ for 
even and $9m_\pi^2$ for odd G parity $N\bar{N}$ channels), the 
phase $\delta(s)$ of the FF defined as 
$G(s)=|G(s)|e^{i\delta(s)}$ is given by the principal value integral
$$
\delta(s) = -{{s\sqrt{s - s_o}}\over{\pi}}
\Pr\int_{s_o}^\infty{{\ln|G(s^\prime)|ds^\prime}
\over{s^\prime(s^\prime - s)\sqrt{s^\prime - s_o}}} 
\eqno(2)
$$
So in principle the method is very simple: (i) use space-like 
data on the left side of Eq.(1) and solve the integral equation 
to find $\ln|G(s)|$ for $s>s_o$. 
(ii) Having determined the modulus $\ln|G(s)|$, use Eq.(2) 
to compute the phase $\delta(s)$.
\par The above procedure has been unsuccessful in the past as it is
an ``ill-posed'' mathematical problem\cite{1,2}. The result depends on the 
input data in an unstable way and an impossible accuracy is needed 
before one arrives to a stable unique solution.

\begin{figure*}[t]\vspace{-.0cm}
\centering
\includegraphics*[width=75mm,height=75mm]{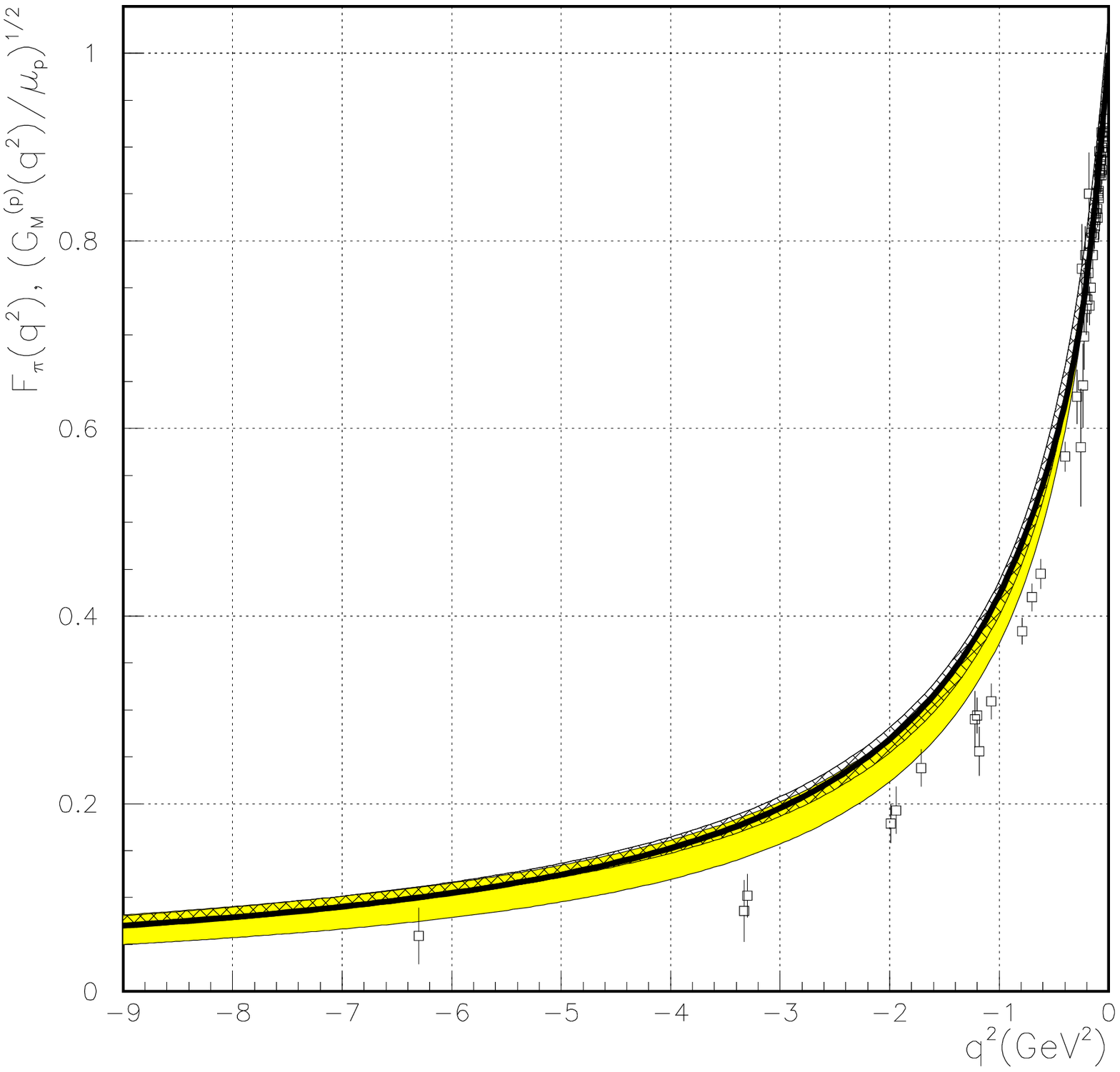}\hspace{1cm}
\includegraphics*[width=75mm,height=75mm]{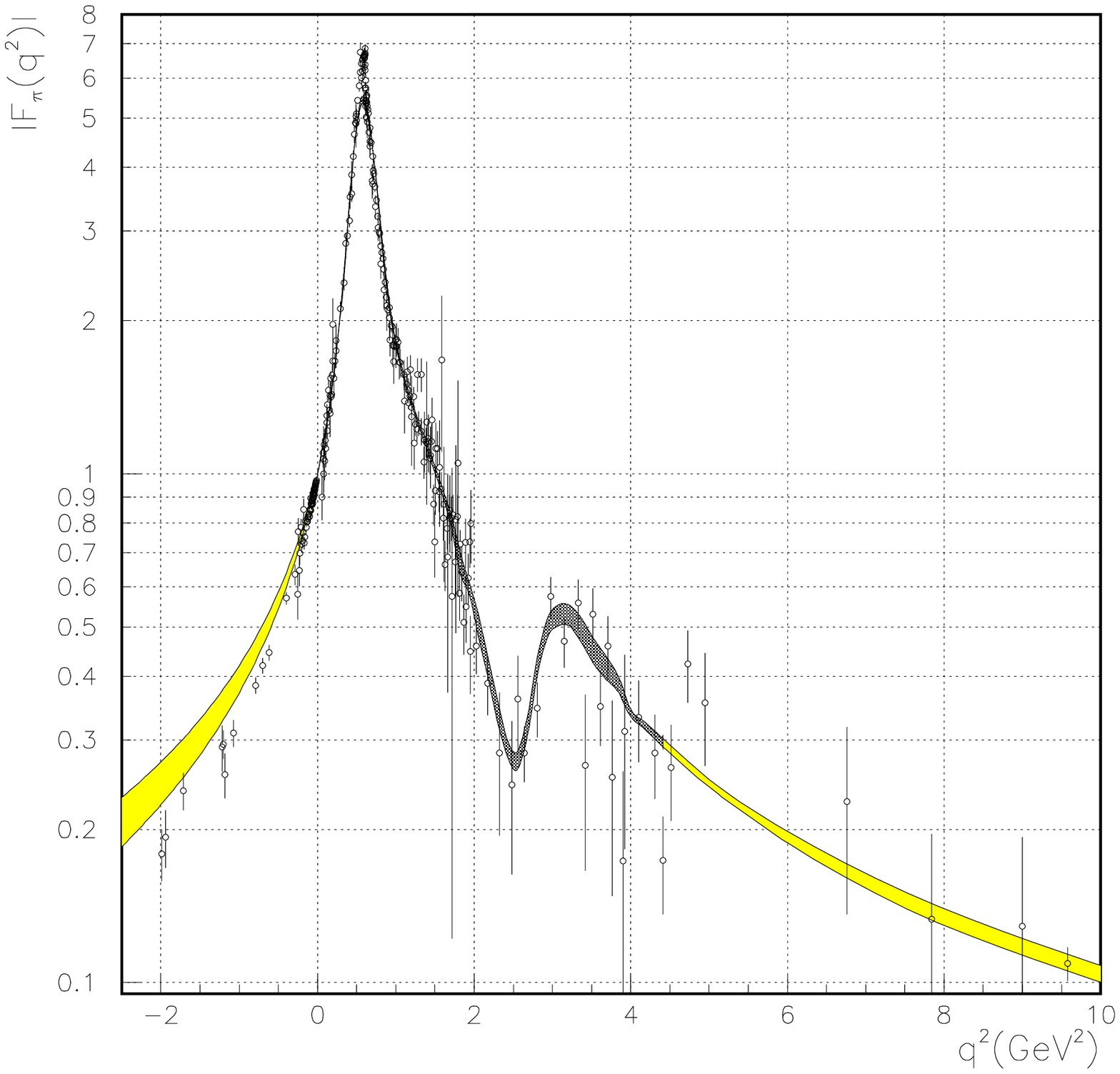}
\begin{minipage}[t]{7.5cm}\vspace{-.5cm}
Figure 1: Pion space-like FF computed via DR without (grid band) 
and with (light band) subtraction, compared with pion FF space-like 
data and square route of proton space-like FF data fit (dark band).
\end{minipage}
\ \hspace{1cm} \
\begin{minipage}[t]{7.5cm}\vspace{-.5cm}
Figure 2: Pion FF computed via DR from time-like data.
\end{minipage}\vspace{-.5cm}
\includegraphics*[width=75mm,height=75mm]{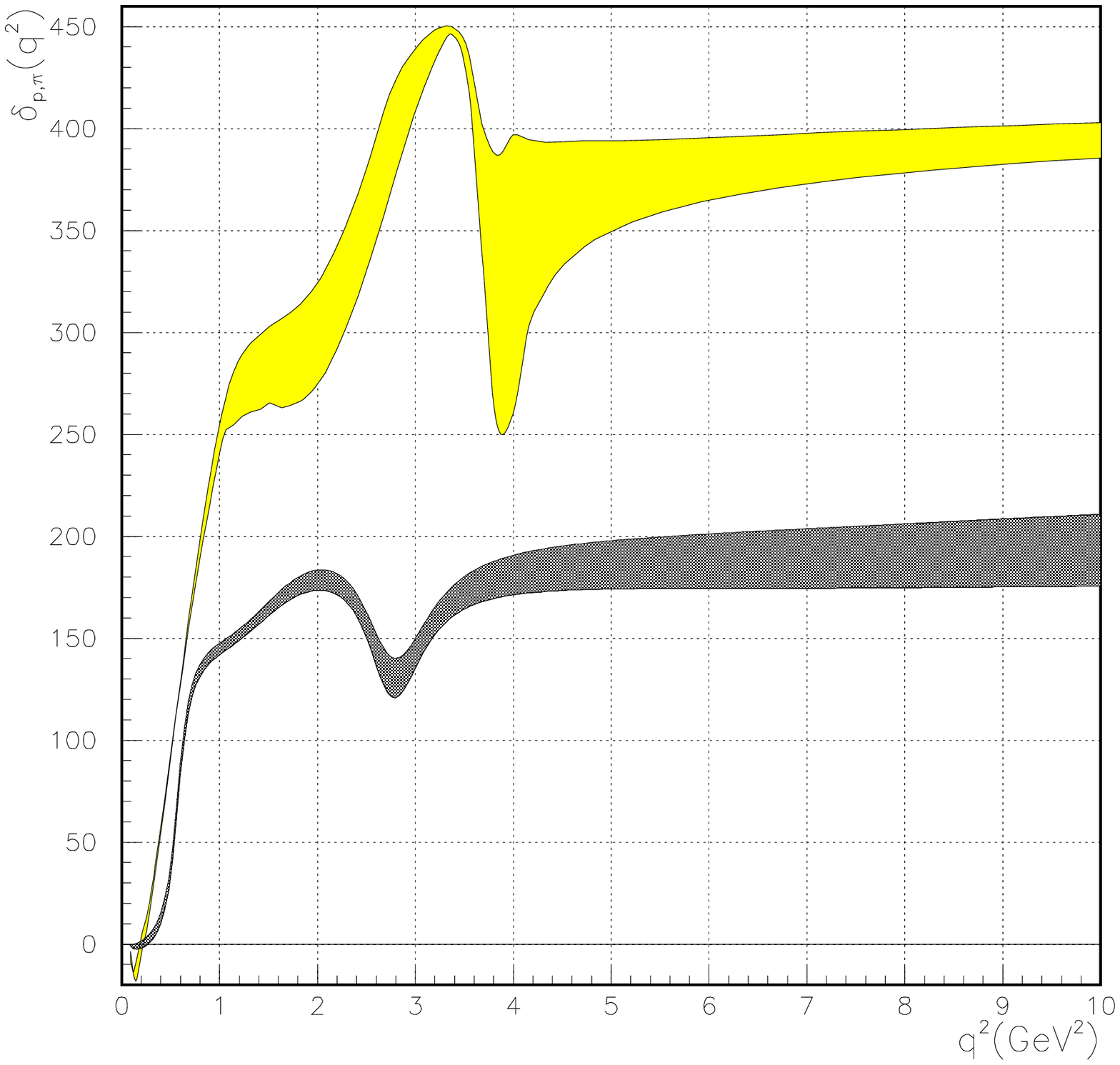}\hspace{1cm}
\includegraphics*[width=75mm,height=75mm]{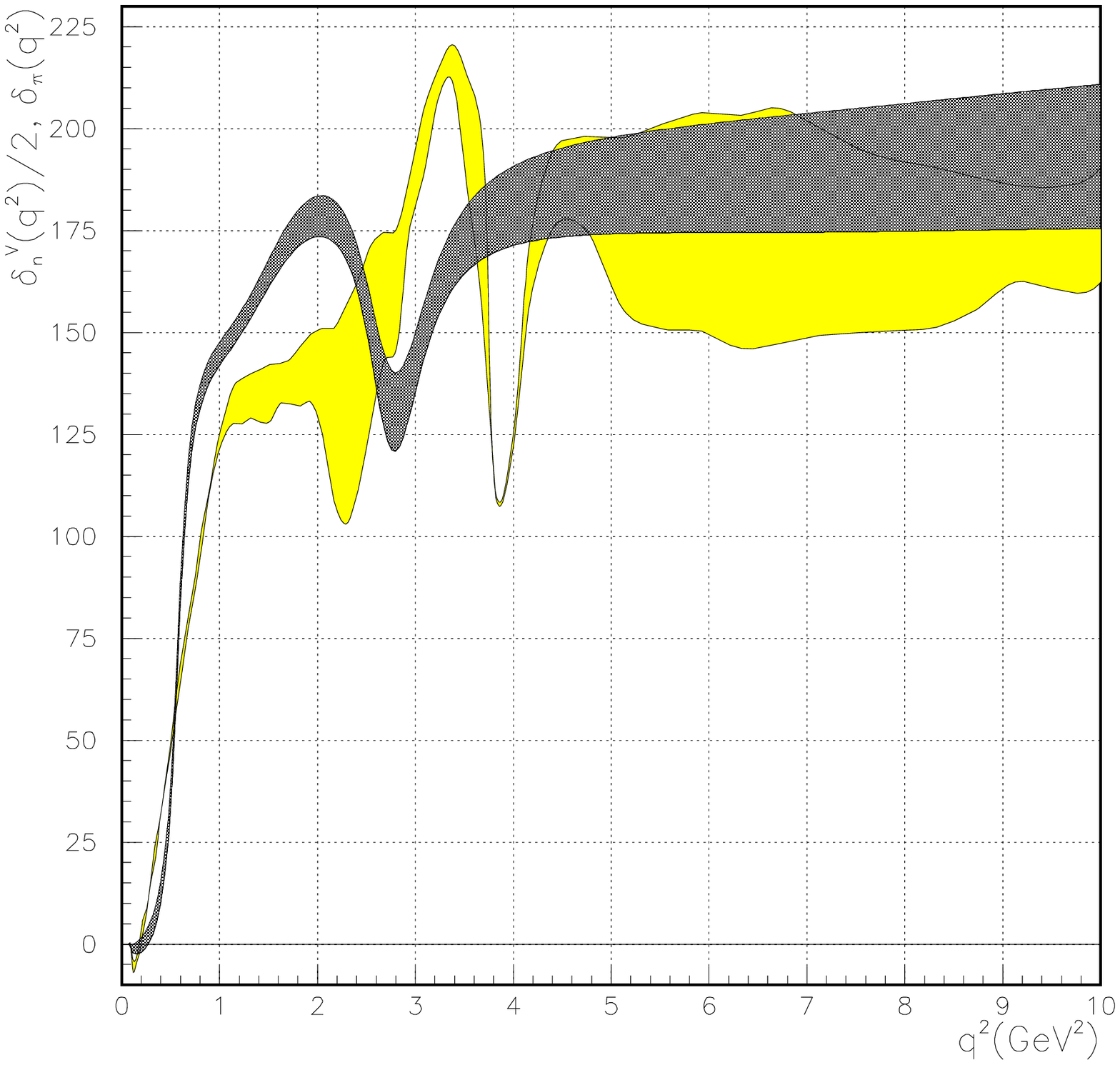}
\begin{minipage}[t]{7.5cm}\vspace{-.5cm}
Figure 3: Phases of proton (light band) and pion (dark band) FF computed 
via DR.
\end{minipage}
\ \hspace{1cm} \
\begin{minipage}[t]{7.5cm}\vspace{-.5cm}
Figure 4: Phase of pion FF (dark band) compared with the half of
nucleon isovector FF (light band).
\end{minipage}\vspace{-0cm}
\end{figure*} 
\begin{figure*}[t]\vspace{0cm}
\centering
\includegraphics*[width=75mm,height=75mm]{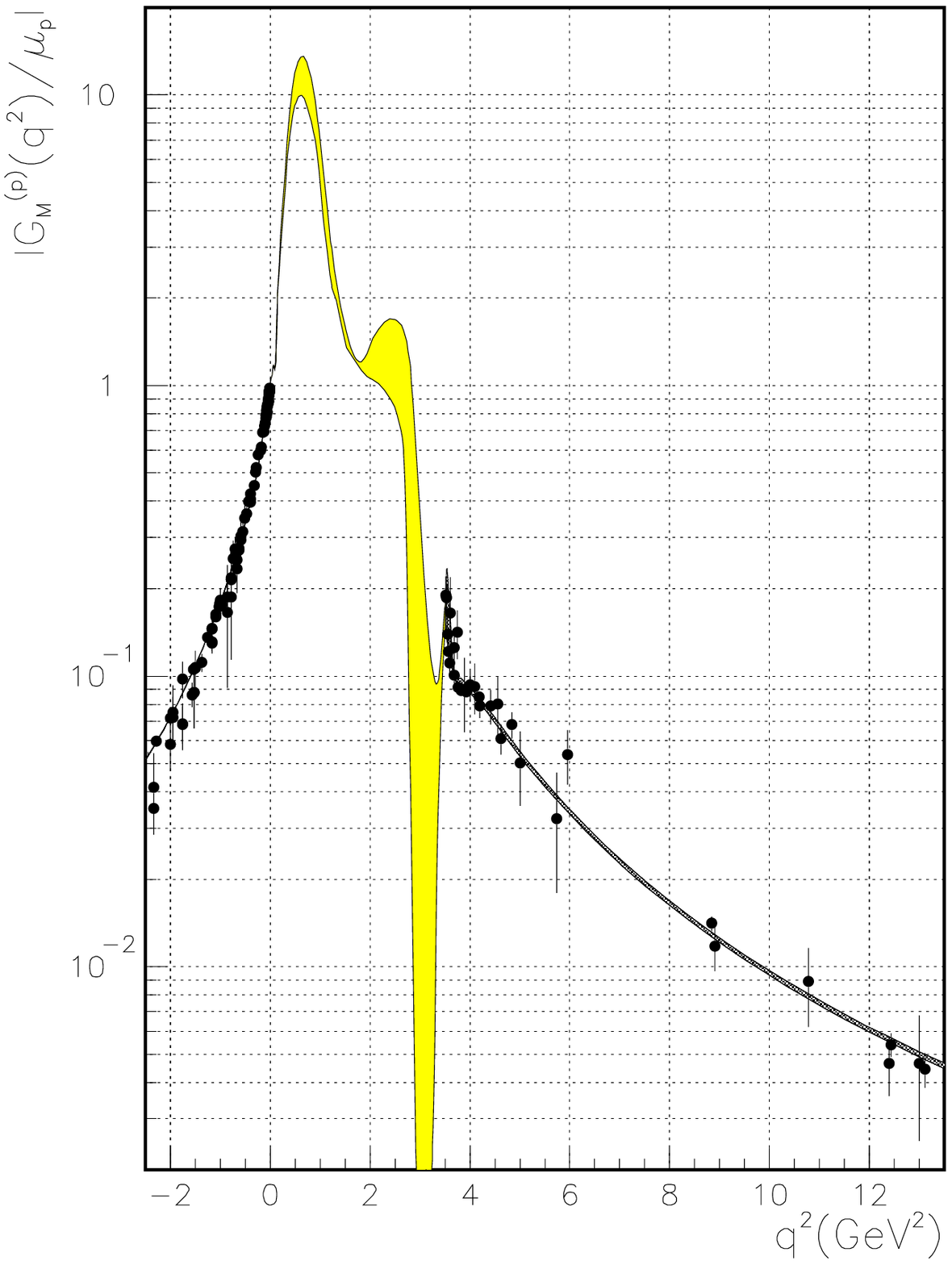}\hspace{1cm}
\includegraphics*[width=75mm,height=75mm]{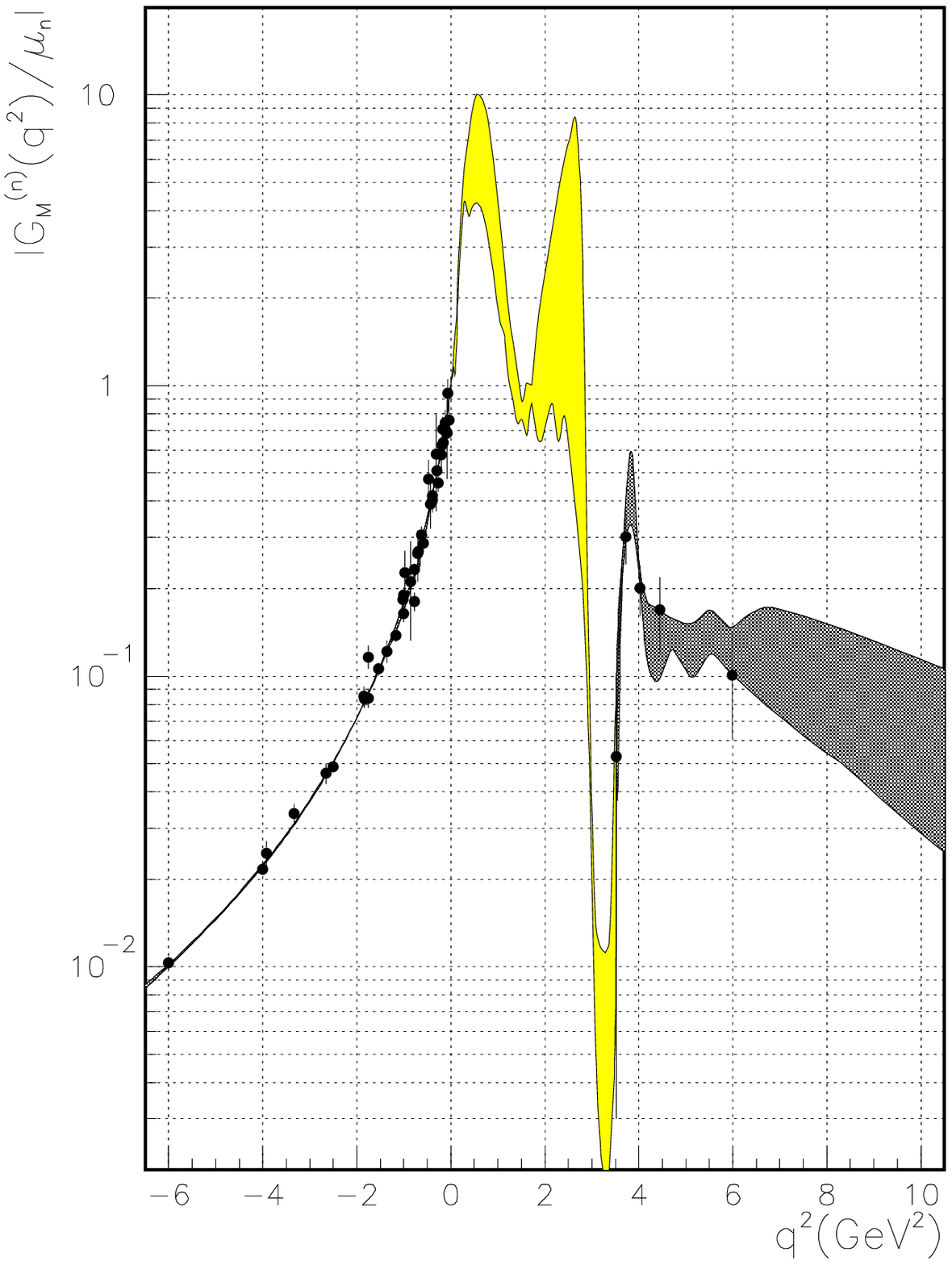}
\begin{minipage}[t]{7.5cm}\vspace{-.5cm}
Figure 5: Proton magnetic FF computed via DR.
\end{minipage}
\ \hspace{1cm} \
\begin{minipage}[t]{7.5cm}\vspace{-.5cm}
Figure 6: Neutron magnetic FF computed via DR.
\end{minipage}\vspace{-.5cm}
\includegraphics*[width=75mm,height=75mm]{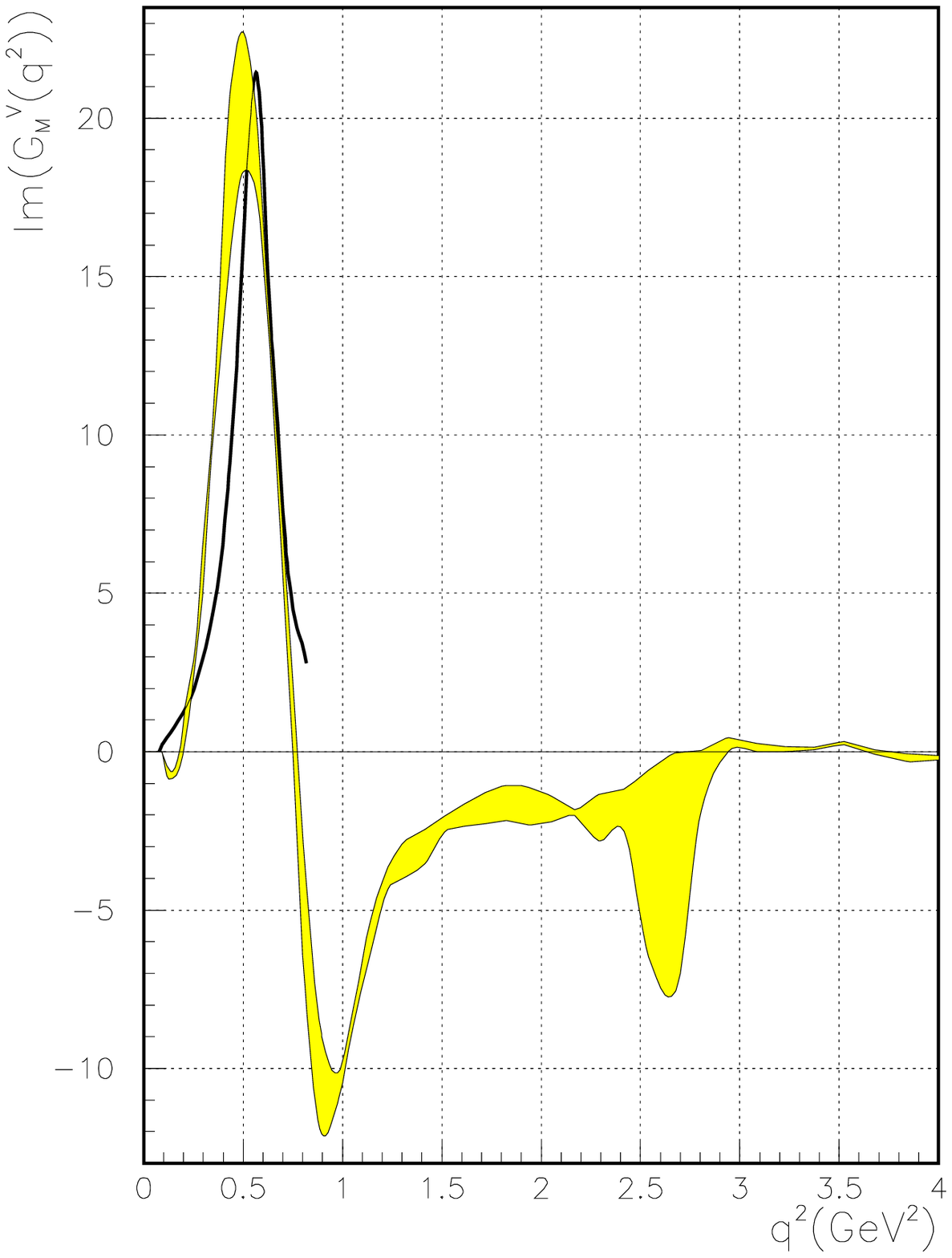}\hspace{1cm}
\includegraphics*[width=75mm,height=75mm]{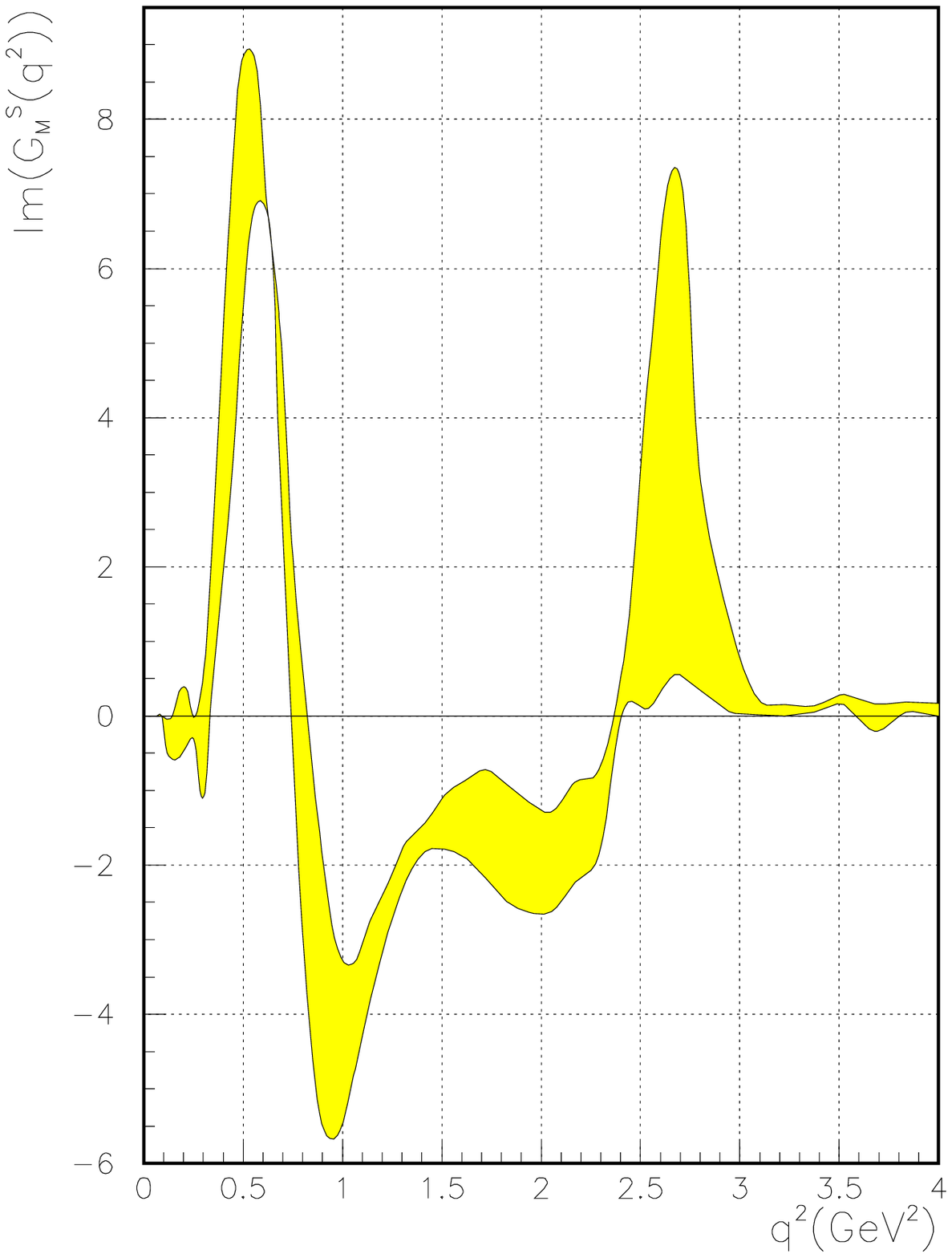}
\begin{minipage}[t]{7.5cm}\vspace{-.5cm}
Figure 7: Imaginary part of the nucleon magnetic isovector FF
computed via DR compared with expectation from unitarity relation.
\end{minipage}
\ \hspace{1cm} \
\begin{minipage}[t]{7.5cm}\vspace{-.5cm}
Figure 8: Imaginary part of the nucleon isoscalar FF computed via
DR.
\end{minipage}\vspace{0cm}
\end{figure*} 
\section{OUR STRATEGY}

A successful method\cite{3} has been developed by splitting the 
time-like region into 2 parts:
\par\noindent (i) \underbar{Region I}: is the unknown unphysical
region $[s_o,4m_N^2]$ for which the FF is to be determined.
\par\noindent (ii) \underbar{Region II}: consists of the physically
accessible time-like region $s>4m_N^2$, for which data exist and 
quite accurate asymptotic estimates are available.\par 
        With the above breakup, the unknown part of the integral 
equation is reduced to the (small) region I, which is amenable to a 
finite matrix analysis with some technical refinements described below
in brief (the details of the developed procedure may be found in 
references \cite{3} and \cite{4}).\par
        An integral equation of the first kind, linear in the unknown
$\ln|G|$, can be derived
$$
\ln G(t) -I(t) =  {{t\sqrt{s_o - t}}\over{\pi}}
\int_{s_o}^{4m_N^2}\!\!\!\! {{\ln|G(s)|ds}\over{s(s - t)\sqrt{s - s_o}}}, 
\eqno(3)
$$
where
$$
I(t) = {{t\sqrt{s_o - t}}\over{\pi}}
\int_{4m_N^2}^\infty{{\ln|G(s)|ds}\over{s(s - t)\sqrt{s - s_o}}}, 
 \eqno(4)
$$
is a ``known'' quantity since it can be calculated directly from  
experimental data in the time-like region with some recipe to 
extrapolate them to very high $t$ values.\par 
        To avoid instabilities usually met in solving first kind integrals
such as Eq.(3), we impose a regularization scheme with smoothness:
\par\noindent
$\bullet$\ 
$I(t)$ is calculated using time-like data through a rational, 
smooth function with the expected asymptotic behavior. The subtraction
at $s=0$ helps in diminishing the impact which the asymptotic 
behavior has on the results.
\par\noindent 
$\bullet$\ 
There is a steep spike near the $N\bar{N}$ 
threshold. To avoid any ensuing instabilities, the upper limit  
of the unphysical region has been raised to $s_2=4m_N^2+\Delta$
where $\Delta\approx 0.5 GeV^2$ and continuity is imposed there. 
A new DR is constructed for the region 
($4m_N^2,s_2 $)\cite{3,4}.\\
\noindent
$\bullet$ Our regularization consists in requiring the local curvature 
of the FF in the unphysical region, $R_2=\int_{s_o}^{s_2}
\big{(} {{d^2|G(s)|}\over{ds^2}}\big{)}^2 ds$ to be limited. 
Instead of the second derivative of $\ln|G|$, as is standard\cite{1,2}, we 
employ the second derivative of $|G|$ for this purpose. The reason is 
that fluctuations in $|G|$ are important only when $|G|$ is 
large, while  $\ln|G|$ fluctuations would be large also when 
$|G|$ is small.\\
\noindent 
$\bullet$ Eq.(3) is then linearized by transforming the integrals into sums 
over $M=50$ suitable sub intervals in $s$, with their widths 
increasing with $s$. This introduces further smoothness, by
effectively integrating over any structure with a narrower half width.
\par\noindent $\bullet$\ The minimize the integral
$$
R_o = \sum_{i = 1}^{L} \Bigg[ \sum_{j = 1}^{M} F_j
{{t_i\sqrt{s_o - t_i}}\over{\pi}}\int_{s_j}^{s_{j+1}}\!\!\!\!\!\!
{{ds}\over{s(s - t_i)\sqrt{s - s_o}}} + \Bigg.
$$
$$
+\Bigg.I(t_i)-
\ln G(t_i) \Bigg]^2, \eqno(5)
$$
where $F_j=\ln|G[(s_{j + 1}+ s_j)/2]|$ is calculated in the
middle of the $j$ th sub interval. $t_i$, with $i  =\
1,\cdots,L$, correspond to the experimental points available in 
he space-like region. 
\par\noindent $\bullet$\ Finally, we have
$$
{\cal R}_{total} = R_o + \tau^6 R_2 + C
$$
The ``dumping parameter'' $\tau$ has to be
chosen by trial and error: if it is set too large, it will not 
respond to sharp structures, while unstable solutions will result 
if it is set too low.
\par\noindent$\bullet$\ The uncertainties in the solution of Eqs.(2)
and (3), due to experimental errors, were estimated by simulating
new space- and time-like data according to the quoted errors and
then solving the DR for each simulated set.
\section{TEST OF THE REGULARIZATION METHOD}

To test the entire procedure and also to get a suitable range for
the parameter $\tau$, we computed the space-like pion FF using 
time-like data. In the time-like region, this FF is known 
up to the $J/\Psi$ mass and at higher $Q^2$, it was extrapolated 
using first order QCD\cite{5}. In Fig.(1), comparison is made with
the measured (low $Q^2$) space-like data. (Higher space like $Q^2$
data points are through extrapolations from pion electroproduction 
data and thus may have systematic errors). We have also made other
tests[3,4] obtaining good agreement with the $\rho$ peak, the $\rho$ 
width and also a dip at $1.6\ GeV^2$ for $\tau\approx m_\pi$ 
Fig.(2). The phase of the pion FF approaches just above $2\ GeV$ to 
its expected asymptotic value of $180$ degrees Fig.(3,4).

\section{RESULTS FOR THE NUCLEON FF}

Below we summarize some salient features
of our findings:\par\noindent
$\bullet$\ For the first time, resonant structures have been 
generated from ``smooth'' inputs Fig.(5,6,7,8). The method is stable and reliable.
\par\noindent $\bullet$\ The combined ($\rho\ +\ \omega $) peaks and
the $\rho'(1600)$ are generated at the right mass. However, the
$\rho$ peak is much broader. Earlier analyses\cite{6} had also found 
a similar discrepancy. 
\par\noindent $\bullet$\ No $\Phi$ signal is visible thus signaling
a very small $s\bar{s}$ content in the nucleon.
\par\noindent $\bullet$\ Phases for the nucleon are consistent with
expectations: $\delta_N\ \rightarrow\ 360^o$ within the error bands Fig.(3).
\par\noindent $\bullet$\ There is an interference pattern near
threshold  ($ M\ \approx 1.88\ GeV$) which may be related to 
baryonium Fig.(5,6).                                    
\par\noindent $\bullet$\ $\mbox{Im}G_M^{(V)}$ changes sign once
and $\mbox{Im}G_M^{(S)}$ appears to change sign twice and
various superconvergence sum rules are all obeyed by our FF's
Fig.(7,8).
By way of comparison, our analysis strongly indicates that 
$\mbox{Im}F_\pi$ does not change sign. Neglecting logaritmic factors
(which we can not resolve), this would suggest (for power law
behavior) that
$$
  |F_\pi(s)| \rightarrow |s|^{-1 + \epsilon}\ \mbox{as}\ |s| \rightarrow
\infty. \ \ (\epsilon\ > 0).
$$
\section{CONCLUSIONS}

Nucleon time-like magnetic FF have been obtained in an 
almost model independent way by means of DR for $\ln G(q^2)$,
using a regularization method in conjunction with space and time-like
data. Resonances have been found consistent with the $\rho(770)$
and $\rho'(1600)$ masses. However, a very large $\rho$ width is
obtained - as in previous DR analyses. Further work is in progress
to understand the sources of discrepancies as well as the
relationship of our results with other DR analyses\cite{7}.\\
Other applications of this strategy have been discussed at this conference
by Y. Srivastava (see contribution T19).

\end{document}